\documentclass[11pt,twoside]{article}  
\usepackage{adassconf}

\begin{document}   

%
%
%

\paperID{P3.23}

%
%
%
%

\title{Comprehensive Metadata Query Interface for Heterogeneous Data Archives Based on Open Source PostgreSQL ORDBMS}
\titlemark{Comprehensive Metadata Querying with PostgreSQL}

%
%
%

\author{Ivan Zolotukhin \altaffilmark{1}}
\author{Nikolay Samokhvalov \altaffilmark{2}}
\author{Francois Bonnarel \altaffilmark{3}}
\author{Igor Chilingarian \altaffilmark{1,4}}

\altaffiltext{1}{Sternberg Astronomical Institute, Moscow State University,
13 Universitetski prospect, Moscow, 119992, Russia}
\altaffiltext{2}{Institute for System Programming of RAS, 25 B. Kommunisticheskaya, Moscow, 109004, Russia}
\altaffiltext{3}{Centre de Donn\'ees Astronomiques de Starsbourg, Observatoire de Starsbourg; CNRS, UMR 7550; 
Universit\'e Louis Pasteur, Starsbourg, France}
\altaffiltext{4}{Centre de Recherche Astronomique de Lyon, Observatoire de Lyon; CNRS, UMR 5574; Universit\'e Claude Bernard Lyon-1;
Ecole Normale Sup\'erieure de Lyon, Lyon, France}

%
%

\contact{Ivan Zolotukhin}
\email{iz@sai.msu.ru}

%
%
%
%
%

\paindex{Zolotukhin, I.}
\aindex{Chilingarian, I.}     
\aindex{Bonnarel, F.}
\aindex{Samokhvalov, N.}

%
%

\authormark{Zolotukhin et al.}

%
%



\begin{abstract}          
We use PostgreSQL DBMS for storing XML metadata, described by the
IVOA Characterisation Data Model. Initial XML type support in the
PostgreSQL has recently been implemented. We make heavy use of this
feature in order to provide comprehensive search over Characterisation
metadata tree. We built a prototype of the Characterisation metadata
query service, implementing two access methods: (1) HTTP-GET/POST
based interface implements almost direct translation of the query
parameter name into XPath of the data model element in the XML
serialisation; (2) Web-Service based interface to receive XQuery which
is also directly translated into XPath. This service will be used in
the ASPID-SR archive, containing science-ready data obtained with the
Russian 6-m telescope.
\end{abstract}

%
%

\section{Introduction}
Storing and querying structured metadata is an important point for building
astronomical archives containing heterogeneous datasets coming from various
telescopes and/or instruments. IVOA Characterisation Data Model (McDowell
et al., in prep.) allows to describe a position of any astronomical dataset
in the multidimensional space of physical parameters. Thus, building query
interface on top of the characterisation metadata, will provide enough
capabilities for elaborated queries often needed for sophisticated
scientific usage of the resource.

\section{Implementation}
Usage of mature and freely available DBMS engine as a backend for the
query interface is of a big importance for further development of
working prototypes implementing Characterisation DM, since relational
database server can solve many data structure and manipulation
problems one may encounter when deploying data archive in a
consistent way.

With the advent of native XML support in relational database engines it all
become possible. We use open source PostgreSQL DBMS for storing and
querying characterisation metadata.

Initial XML type support in the PostgreSQL has recently been implemented by
NS in a frame of ''Google Summer of Code 2006''. We make heavy use of this
feature in order to provide comprehensive search over Characterisation
metadata tree.

\begin{figure}
\epsscale{0.9}
\plotone{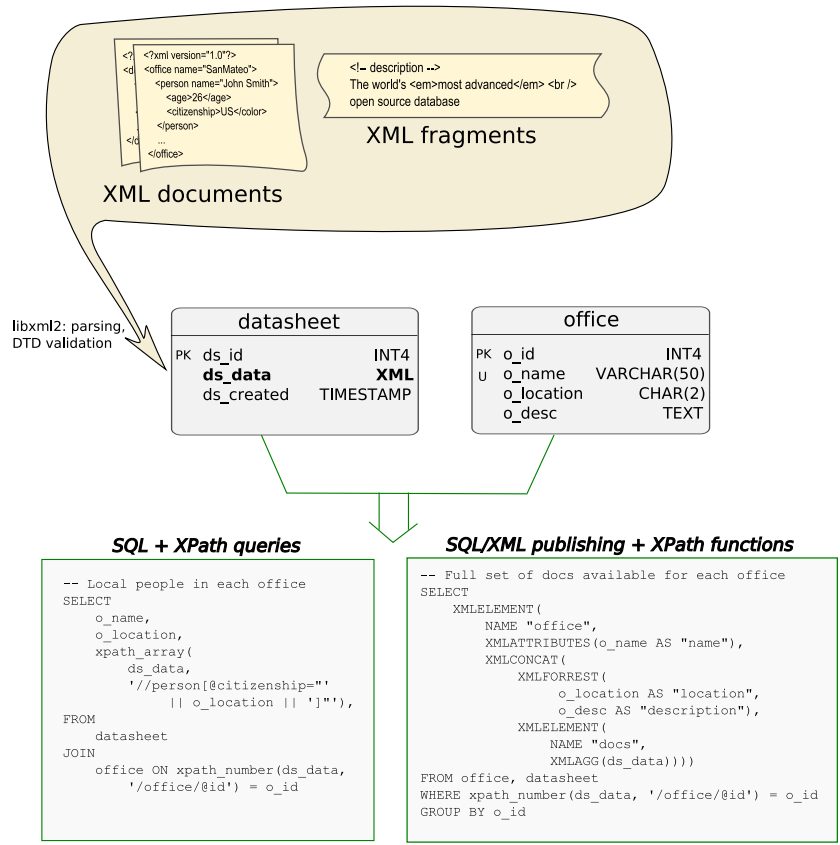}
\caption{Incorporation of XPath in SQL queries
\label{P3.23_fig1}}
\end{figure}

Figure~\ref{P3.23_fig1} demonstrates the concepts used to store and query XML
structures, and shows how XPath expressions can be incorporated in SQL
queries.

We built a prototype of the Characterisation metadata query service,
implementing two access methods:
\begin{enumerate}
\item HTTP-GET/POST based interface implements almost direct
translation of the query parameter name into XPath of the data model
element, then incorporated into extended SQL/XML queries to XML and
relational data being stored in DBMS. This concept allows to
distinguish axes by some specific properties, for instance by their
UCDs, and provides ability of putting constraints on the query result.
\item Web-Service based interface to receive XQuery which is also directly
translated into SQL/XML statements. This method aims in dealing with
ADQL-like queries in the future.
\end{enumerate}

\section{Perspectives}
Presently XML type in PostgreSQL is developed for manipulation
abilities rather than a specific storage engine, meaning that it is
built on top of VARCHAR as initial storage implicit type. This somehow limits
performance of queries since one is confined only to use
functional indices based on result of XPath expression evaluation.

Nearest plans of XML type development include very important features that
will comprise:
\begin{itemize}
\item XMLQUERY -- standardized by SQL:2006 way of integration
of XQuery capabilities with other essential relational
functionality. Joins of XML and relational data in one expression.
\item Proper design at physical level (data structure, comprehensive
  indices system and Generalized Search Tree usage, etc). This will
  lead to fast XQuery and XPath evaluation.
\end{itemize}

Thus, all performance penalties for this combination of relational and
XML approaches will soon be minimized dramatically.

This opens very broad perspectives for building similar systems making
use of some XML-based data models since one can achieve all that
flexibility and comprehensiveness right inside DBMS.

\acknowledgments
We are very grateful to the support given by the organizing committee of
ADASS, essential for attending this exciting conference. Travel of IZ is
supported via RFBR grant \#06-02-27333 and grant of the President of RF for
leading scientific school, NSH-5290.2006.2.

\end{document}